\documentclass[twocolumn,preprintnumbers,superscriptaddress,footinbib]{revtex4}
\usepackage{psfrag,epsfig,rotate,bm}
\begin{document}
\title{Three-dimensional magnetic flux-closure patterns in mesoscopic Fe islands}
\author{R. Hertel}
\email{r.hertel@fz-juelich.de}
\affiliation{Institute of Solid State Research (IFF), Research Center J\"ulich, D-52425 J\"ulich, Germany} 
\author{O. Fruchart} 
\author{S. Cherifi}
\affiliation{Laboratoire Louis N\'eel, CNRS-UJF-INPG, BP166, F-38042 Grenoble Cedex 9, France}
\author{P.-O. Jubert}
\affiliation{IBM Research, Zurich Research Laboratory, CH-8803 R\"uschlikon, Switzerland}
\author{S. Heun}
\affiliation{TASC-INFM Laboratory, Area di Ricerca, Basovizza, I-34012 Trieste (TS), Italy} 
\author{A. Locatelli}
\affiliation{Sincrotrone ELETTRA, I-34012 Basovizza, Trieste, Italy}
\author{J. Kirschner}
\affiliation{ Max-Planck--Institute of Microstructure Physics, Weinberg 2, 
06120 Halle, Germany}
\newcommand{\be}{\begin{equation}}
\newcommand{\ee}{\end{equation}}
\newcommand{\p}{\partial}
\newcommand{\heff}{\bm{H}_{\rm eff}}
\newcommand{\dmdt}{\frac{{\rm d}\bm{M}}{{\rm d}t}}
\newcommand{\djdt}{\frac{{\rm d}\bm{J}}{{\rm d}t}}
\begin{abstract}
We have investigated three-dimensional magnetization structures in
numerous mesoscopic Fe/Mo(110) islands by means of x-ray magnetic
circular dichroism combined with photoemission electron microscopy
(XMCD-PEEM). The particles are epitaxial islands with an elongated
hexagonal shape with length of up to 2.5 $\mu$m and  
thickness of up to 250 nm. The XMCD-PEEM studies reveal 
asymmetric magnetization distributions at the surface of these
particles. Micromagnetic simulations are in excellent agreement with
the observed magnetic structures and provide information on the
internal structure of the magnetization which is not accessible in the
experiment. It is shown that the magnetization is influenced mostly by
the particle size and thickness rather than by the details of its
shape. Hence, these hexagonal samples can be regarded as model systems
for the study of the magnetization in thick, mesoscopic ferromagnets.  
\end{abstract}
\maketitle
\section{Introduction} 
The spatial confinement of the magnetization in a 
ferromagnet of mesoscopic or nanometric size can have a  
dramatic impact on both the magnetic properties 
and the magnetization distribution 
\cite{Martin03,Cowburn99,Kirk97,Gomez99,Hubert98}. 
Besides the technological importance of understanding and controlling
such finite-size effects in magnetic nanoparticles that can be used
for magneto-electronic devices, the physics of the magnetization in
confined structures is a beautiful and exciting research topic
\footnote{The numerous images of magnetic domains in the
  famous textbook by A.~Hubert and R.~Sch\"afer \cite{Hubert98} may
  give the reader an impression of the beauty of magnetic domain
  patterns.}.  
A precise theoretical description of the magnetization distribution 
in particles of a size between about 100 nm and a few microns 
can be obtained in the framework of micromagnetism
 \cite{Brown63,Aharoni96_2nd}. Generally, the micromagnetic equations can only
 be solved numerically. Mesoscopic magnetic
particles are too small for a meaningful description in terms of
macroscopic  material constants like the magnetic susceptibility.
 On the other hand, the simple
macrospin model -- which could be solved analytically \cite{Stoner48} -- 
can only be applied for very small ($\lesssim$10 nm) magnetic
particles \cite{Wernsdorfer97}. It loses its validity in larger ferromagnets, 
where the magnetization distribution is strongly inhomogeneous.  

The subtle size-dependence of the magnetization \cite{Yamasaki03,Hertel02c,Usov01}
is particularly important when the particle size is comparable to the
so-called magnetic exchange length \cite{Hubert98,Kroni62},
{\em i.e.} a material-dependent length scale that describes the order of
magnitude for the extension of magnetic inhomogeneities like domain
walls or magnetic vortices. Besides the size, also the particle  
shape is known to have a strong influence  on the magnetic properties
of nanostructures. Numerous studies on magnetic particles of 
different  geometries, like rings \cite{Rothman01}, rectangles
\cite{Gomez99}, nanowires \cite{Nielsch2001} and various other geometries 
\cite{Park03,Cowburn99,Kirk01} have been reported 
in the last
years. Apart from a few exceptions where three-dimensional magnetic
structures have been investigated numerically ({\em e.g.}
\cite{Hertel99b,Rave98,Ha03a}), experimentally \cite{Hehn96} or both
\cite{Jubert03}, most studies on magnetic nanostructures that have
been published over the last years have focussed on particles in which
{\em at least one} dimension is so small that    
the magnetization in the sample is either two-dimensional (thin-film
elements) or one-dimensional (thin nanowires). 
Therefore, compared to thin film elements, where the influence of the
particle shape on the magnetic properties has been amply studied, not
much is known about the effect of spatial confinement in
three-dimensional magnetic particles.  

Similarly, several types of magnetic domain walls 
\cite{Bloch32,Neel55,Moon59,Hubert69} have been
investigated thoroughly in extended magnetic films
\cite{Hubert98}. In these studies, the sample was essentially confined in one
dimension (the film thickness). Recently, the influence of geometric
confinement on domain walls has been studied in systems with
two-dimensional \cite{Jubert04} and one-dimensional \cite{Bruno99}
magnetization, and different types of head-to-head domain
walls have been predicted  \cite{McMichael97b} and observed
\cite{Klaui04} in  thin
magnetic strips or rings of different width. However, the effect of a
mesoscopic spatial confinement of {\em three-dimensional} magnetic
domain walls has hardly been investigated up to now.

In this paper we report on three-dimensional magnetization
distributions in mesoscopic ferromagnets. As a model system, we have
analyzed several monocrystalline Fe islands of different size 
(up to ca.~2.5 $\mu$m length and up to ca.~250 nm thickness). 
These particles are small
enough to display pronounced finite-size effects of the magnetization,
while they are sufficiently large and thick to sustain inhomogeneous,
three-dimensional arrangements of the magnetization. The experimental part of
this study is performed by means of X-ray magnetic circular 
dichroism with photoemission electron microscopy (XMCD-PEEM). The
direct comparison of high-resolution experimental observations with
highly accurate micromagnetic computer simulations allows for an
unambiguous interpretation of some complicated, unexpected magnetization
distributions.  This analysis leads us to the
conclusion that the Landau pattern, which is well-known from
magnetic thin-film elements, is merely a simple variant of a general,
more complex magnetization structure.

\section{sample preparation}
\begin{figure}
\centerline{\epsfig{file=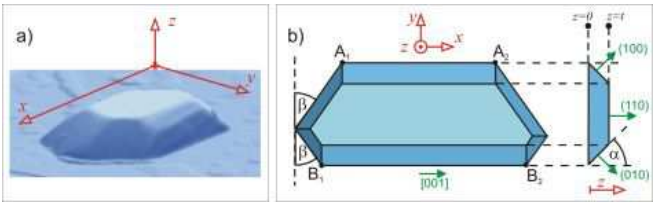,width=\linewidth}}
\caption{\label{geom}(color online) (a) Atomic force microscopy of a Fe island, about 1$\mu$m long. (b)
  Schematic presentation of a typical particle geometry. The angles
  $\alpha=45^\circ$ and $\beta=35.26^\circ$ are fixed parameters that
  result from the monocrystalline structure of the islands. The
  shape is uniquely defined by the points $A_1$, $A_2$, $B_1$,
  $B_2$ and the thickness $t$. Note that this set of parameters is
  subject to various constraints. For instance, the edges ($A_1,A_2$)
  and ($B_1,B_2$) are parallel to each other (the
  $x$-direction). }
\end{figure}

Mesoscopic monocrystalline Fe islands have been grown on atomically flat Mo(110) buffer layers of
thickness 10 nm \cite{Fruchart98}, deposited on Al$_2$O$_3(11-20)$. The growth is performed with
pulsed laser deposition (PLD) in ultra-high vacuum at 850 K. Under these conditions, Fe islands
are formed by means of the Stranski-Krastanov growth mode. A cover layer of Mo[three atomic
layers], followed by Al[3nm] has been added to protect the samples from oxidation. A more detailed
description of the fabrication of self-assembled Fe/Mo(110) islands is reported elsewhere
\cite{Jubert01,Fruchart04}. The islands have a hexagonal shape, a flat top and inclined lateral
facets, all atomically flat, see Fig.~\ref{geom}a).

If the growth proceeded under thermodynamic equilibrium, the shape of all islands would be
expected to be identical, and described by the Wulff-Kaischev theorem (see \cite{Mueller00} and references therein). We indeed
observe that the islands share a common set of features: the angle of inclination of the lateral
facets, the horizontal top facet, as well as the direction of all edges. However, the shape of the
particles is generally not symmetric and not uniform. The vertical and in-plane aspect ratios
slightly differ from one to another. This may arise from kinetic effects, the influence of atomic
steps on the Mo(110) surface or residual strain. The size ranges between about 500 nm to 2500 nm
in length and about 50 to 250 nm in thickness. Our observations suggest that the shape of the
particles is well reproduced by taking into account only \{001\} and \{110\} facets. The
examination of the constraints related to these facets  leads to an idealized geometrical
construction scheme for the shape of these islands as described in Fig~\ref{geom}. Using
this scheme, each island is uniquely described by the four points $A_1$,$A_2$,$B_1$,$B_2$ and the
thickness $t$.
Approximate values for the thickness and the
  coordinates of these four points have been extracted from electron
  microscopy images to construct models that were used as input for the
  numerical simulations.
The density of the islands on the substrate is low enough, so that
generally speaking, the magnetostatic coupling between the samples is
negligible. Obviously, a few
 exceptional cases can also be found, where two islands have
formed in close vicinity.

These self-assembled Fe islands can be regarded as a
model system for the study of magnetization distributions in
three-dimensional, mesoscopic particles. Due to their high structural
quality the Fe/Mo(110) islands are particularly suited for this
investigation.
A comparable quality would be diffult to achieve with lithographically
fabricated samples, especially for particles of such elevated
thickness. The well-defined shape and structure of the
self-assembled islands ensures that the magnetization distribution in the
particle is governed by the spatial confinement, and not by  roughness
effects or by other structural inhomogeneities. However, the
consideration of these islands as  {\em model systems} for the
magnetization in thick, structured magnetic
particles is only valid if the details of the particle shape (e.g. the
inclination angle of the facets or the precise hexagonal shape) are
not of decisive importance for the resulting magnetic structure. Else, each
particle would have its own characteristic magnetic structure
depending on its shape, thus precluding an extraction of
valuable information of general validity. As will be shown in section
\ref{generalized}, computer simulations demonstrate that
the aspect ratio, the size and the thickness are the most important
parameters, while details of the shape  play only a minor role concerning
the overall magnetization structure.

\section{Experimental setup}

\begin{figure}
\centerline{\epsfig{file=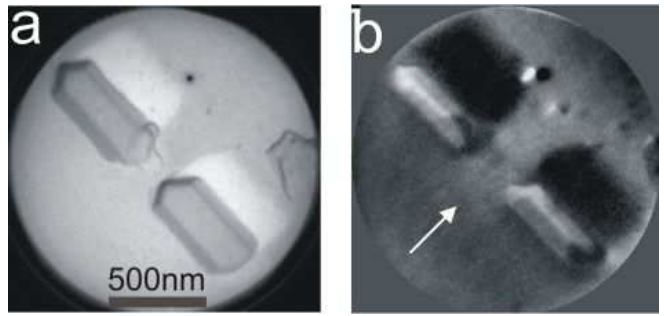, width=\linewidth}}
\caption{\label{example}Example of an experimental observation of the
  particle shape and the magnetization structure. (a): Two Fe islands,
  incidentally located next to each other, imaged with LEEM. 
  (b): The XMCD-PEEM image
  displays the in-plane magnetization component parallel to the incident
  beam. The beam direction is sketched by the white arrow. Due to
  grazing incidence of the photons (74$^\circ$ with respect to the
  plane normal) the back side of the island is shadowed, thus it
  appears dark in the XMCD-PEEM image.
}
\end{figure}

The experimental investigation of the magnetic structures was
performed with XMCD-PEEM \cite{Stoehr93}. The principle of XMCD-PEEM is the 
spin- and helicity-dependent cross-section for photoelectron emission
with circularly  polarized X-rays. 
The lateral resolution of this technique is now better than 30 nm
\cite{Locatelli02}, so that detailed images of the domain structure in
mesoscopic elements can be obtained.
The experiments have been carried out at the
Nanospectroscopy beamline of the ELETTRA synchrotron radiation
facility \cite{Locatelli03}. The combination of low-energy electron
microscopy (LEEM) and PEEM at 
ELETTRA makes it possible to image both the morphology and the
magnetic structure with the same instrument. In the XMCD-PEEM mode,
the samples have been 
illuminated with monochromatic, elliptically polarized
X-rays in a micro-spot of 20 $\times$ 5\,$\mu$m$^2$ (in the horizontal
and vertical direction, respectively). The local
photoelectron current density was imaged with 
PEEM. The photoelectron  density is proportional to
the magneto-dichroic signal $\bm{\sigma}\cdot\bm{M}$, where 
$\bm{\sigma}$ is a vector that indicates the direction of the incident
light and its helicity and $\bm{M}$ is the local magnetization
vector. 
Areas with $\bm{M}$ parallel or antiparallel to $\bm{\sigma}$ have
different secondary electron yields, which leads to a bright and dark
contrast. 
In most cases, we have chosen an irradiation direction $\bm{\sigma}$ perpendicular
to the magnetization direction $\bm{M}$ in the major
domains. Although in this case the major domains show no magnetic
contrast, this choice of the beam direction is advantageous because the
particularly interesting regions that separate the domains, {\em i.e.}
the domain walls, display a strong magneto-dichroic signal. 
The imaging of the domain walls requires a very high spatial
resolution. In the energy range of the Fe 2p$_{3/2}$ (L$_3$) absortion edge, the
photon flux in the illuminated area is about $10^{11}$ photons per
second, which allows for a fast acquisition of single images, in the
range from 10\,s to 30\,s for images with a field of view of $2.5\,\mu$m diameter.
The best lateral resolution in XMCD-PEEM mode has been achieved by
recording series of about 30 images for each photon beam helicity. The
images are post-processed using a self-correlation algorithm
\footnote{{\tt http://www.elettra.trieste.it/nanospectroscopy/ userarea/dataanalysis.html}}
to minimize drift effects. Finally, the magnetic contrast is obtained
by subtracting graphically the two accumulated images at opposite
light helicities. The magnetic domain configurations have been
correlated to the precise morphology of the nanostructures using
LEEM. The contrast in LEEM is determined by the surface topography and
its crystalline structure. In addition to the diffraction contrast,
the so-called interference contrast allows one to image surface steps
and thickness gradients in thin films with atomic depth sensitivity
and a lateral resolution of few tens nm. This LEEM interference
contrast allowed us to unambiguously determine the geometry of the
self-organized Fe nanostructures and of their surrounding 45$^\circ$
inclined facets. The quantitative values extracted from the LEEM
images (islands' shape, length, width and height) served as input for
the micromagnetic simulations.
A typical LEEM image and an XMCD-PEEM image of the same sample are
shown in Fig.~\ref{example}(a) and (b), respectively.

\section{Micromagnetics and Numerical Method}
In the framework of micromagnetism the magnetization inside the sample
is represented as a directional field $\bm{M}=\bm{M}(\rm{r})$ with the
constraint $\left|\bm{M}\right|=M_{\rm  s}={\rm const.}$, where
$M_{\rm s}$ is the saturation magnetization. For a description of the 
intrinsic properties of the magnetic material, a set of material constants
is used. Besides the saturation magnetization $M_{\rm  s}$, the
exchange constant $A$ and the anisotropy constant $K$ are required. 
The material parameters are connected with micromagnetic energy
terms. In our case of magnetic particles with cubic anisotropy in
absence of an external magnetic field, the relevant energy terms are
the exchange energy 
\be
E_{\rm exc}=\int\limits_{(V)}A\left[\left(\bm{\nabla}\bm{m}_x\right)^2
+\left(\bm{\nabla}\bm{m}_y\right)^2+\left(\bm{\nabla}\bm{m}_z\right)^2\right]
\,{\rm d}V\quad,
\ee
the cubic anisotropy energy
\be\label{cubnrg}
E_{\rm cub}=\int\limits_{(V)}K_{\rm c}\left(\alpha_1^2\alpha_2^2+
\alpha_1^2\alpha_3^2+\alpha_2^2\alpha_3^2\right)
\,{\rm d}V
\ee
and the stray field energy 
\be\label{strnrg}
E_{\rm stray}=-\frac{\mu_0}{2}\int\limits_{(V)}\bm{M}\cdot\bm{H}_{\rm
  s}\,{\rm d}V.
\ee
We have used $K_{\rm c}=4.8\times10^4$ J/m$^3$, $A=2.0\cdot10^{-11}$J/m
and $M_{\rm s}=1.73\times10^6$ A/m to describe the material of the Fe islands.
The easy axes are oriented along the [001], [100] and [010] directions,
cf.~Fig.~\ref{geom}b). 
In eq.~(\ref{cubnrg}), $\alpha_1$, $\alpha_2$ and $\alpha_3$ are the
directional cosines of the magnetization $\bm{M}$ with respect to the
cubic easy anisotropy axes. 
These material parameters lead to a quality factor $Q=2K/\mu_0M_{\rm
  s}^2\simeq 2.5\times10^{-2}$. A value of $Q\ll1$ indicates that the
material is magnetically soft, i.e. that the arrangement of the
magnetization is primarily driven by the need to minimize the
demagnetizing energy rather than the anisotropy energy. \footnote{Note
  that the $Q$ factor is originally defined for crystals of {\em uniaxial}
  magneto-crystalline anisotropy. A comparison between the maximum stray
  field energy and the maximum uniaxial anisotropy energy leads to the
  argument used here for $Q\ll1$. In the present case of {\em cubic}
  anisotropy, the quantitative interpretation of $Q$ is less rigorous,
but we can still assume that a value $Q\ll1$ corresponds to a
magnetically soft material.} 
 In eq.~(\ref{strnrg}), $\bm{H}_{\rm
  s}=-\bm{\nabla}U$ is the stray field and $U$ is the
magnetic scalar potential. The potential $U$ satisfies Poisson's
equation
$\Delta U=\rho/\mu_0$ inside the sample volume $V$ and the Laplace equation
$\Delta U=0$
outside the sample. The volume charges $\rho=\mu_0\bm{\nabla}\bm{M}$
and the surface charges $\sigma=\bm{n}\cdot\bm{M}$ ($\bm{n}$:
outward surface normal vector) are the sources of the stray field. 
 The reduction or avoidance of these charges lowers the stray field 
energy \cite{Aharoni96_2nd}. For the inward limit $U_{\rm i}$ and the outward limit $U_{\rm o}$ of 
the potential at the surface, the Neumann boundary conditions at the surface $\p V$ read
\be
\left.\frac{\p U_{\rm i}}{\p \bm{n}}\right|_{\p V}-
\left.\frac{\p  U_{\rm o}}{\p \bm{n}}\right|_{\p V}=\sigma\quad.
\ee
In addition, the boundary condition $\lim_{x\to\infty}U(x)=0$ has to be
fulfilled. Details on the numerical calculation of $U$ by means of a
combined boundary element / finite element method are given elsewhere
\cite{Hertel01b}. 

If the material parameters and the anisotropy axes are specified, the
magnetic energy of the sample is uniquely determined by the
magnetization distribution $\bm{M}(\bm{r})$.
In the static equilibrium, the magnetization arranges in a way to
minimize the total energy $E_{\rm tot}=E_{\rm exc}+E_{\rm cub}+E_{\rm
  stray}$. In the computer simulation, the minimization is performed
by using the conjugate gradient method \cite{NumRec}. The minimization
has to be performed under the constraint $\left|\bm{M}\right|=M_{\rm
  s}$, which can be easily ensured by representing the local magnetization
direction with spherical coordinates $\vartheta$ and $\varphi$.

The numerical representation and the micromagnetic modelling is
performed with the finite element method (FEM), using a code developed by
Hertel \cite{Hertel01b}. In the finite element representation, the
sample's volume is subdivided into tetrahedral elements of irregular
size and shape. The magnetization is discretized at the corner points
of the elements (the nodes). A piecewise linear representation of the
magnetization is obtained by means of normalized, linear interpolation
functions that are defined in each element (the so-called shape
functions). The FEM is a particularly powerful and accurate technique
for micromagnetic simulations. It allows the smooth modelling of
particles of general geometry (with curved or inclined surfaces) and
gives the  possibility to apply adaptive mesh refinement techniques.  
These unique features of the FEM
are of essential importance for the
present study where large samples with a complicated shape are
modelled.

\section{The Van den Berg scheme}
\begin{figure}
\centerline{\epsfig{file=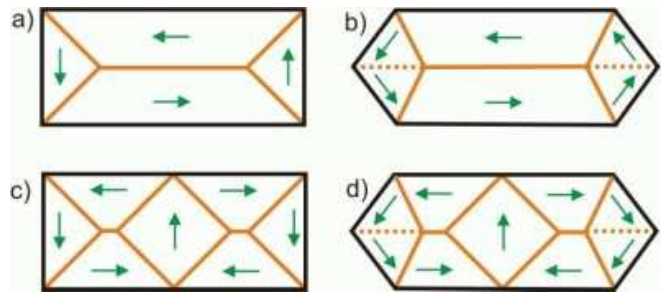,width=\linewidth}}
\caption{\label{VdB}(color online) Simple magnetic flux-closure patterns in thin-film
elements. a), c): Landau und diamond structure in a rectangle,
respectively, b),d): modified Landau and diamond structure,
respectively, in a hexagonal particle. In the hexagonal element, the
closure domains are subdivided in two regions separated by an
additional domain wall (dashed line). } 
\end{figure}
In thin-film elements, magnetic flux-closure patterns
are relatively simple and well-known. The Landau pattern and the
diamond pattern (cf.~Fig.~\ref{VdB}) are typical examples
thereof. Such flux-closure patterns are determined by the
particle's boundary and the tendency to avoid magnetic surface
charges. In sufficiently large soft-magnetic thin-film
elements, the magnetization is oriented in the film plane and is
locally aligned parallel to the particle edges.
This avoidance of magnetic surface charges gives rise to the formation
of magnetic domains.    
The Van den Berg scheme 
\cite{Hubert98,VdBerg86,VdBerg87} is a simple graphical method for
the construction of flux-closure patterns in thin-film elements of
arbitrary shape. This method is derived from a rather complicated
mathematical analysis that assumes an idealized model for
two-dimensional thin film elements with (a) no magnetic anisotropy, 
(b) perfect avoidance of magnetic charges, and (c) domain walls of
vanishing width.
According to this scheme, the center point of each
circle that touches the particle's boundary at least twice is placed
on a domain wall. This construction scheme generally leads to
symmetric domain structures, that, despite the idealizations assumed
in the model, correspond very well to domain
structures as they are obtained in  experiments \cite{Gomez99} and
micromagnetic simulations \cite{Hertel02c,Rave00}. The Van den Berg
scheme was also found to be valid in self-assembled Fe/Mo(110)
elements of moderate thickness ($t\lesssim $60 nm) \cite{Jubert03}.  
In our thicker samples, however, we observe
flux-closure  domain structures that differ significantly from the Van
den Berg scheme. It should be noted that the occurrence of deviations
from the Van den Berg in thick particles is, {\em per se}, not
surprising, since the scheme has been derived using a thin-film 
approximation.  A comparable scheme has not yet been reported for
three-dimensional samples. Analytic studies on the magnetic structure
in three-dimensional magnetic structures have been reported by Arrott
{\em et. al.} \cite{Arrott79,Arrott79b}, which were, however,
restricted to special particle shapes.
In this paper
we report on systematic, qualitative deviations from the Van den Berg
scheme in particles of elevated thickness.  
In order to highlight these differences, we shall refer to the
idealized flux-closure pattern according to Fig.\ref{VdB}a) as the
{\em classical} Landau structure, which will serve us as a
paradigm for magnetic domain structures in thin-film elements. This
term is coined with reference 
to the original work by Landau \cite{Landau35}, where the internal wall
structure had been neglected. The classical Landau structure is a
flux-closure domain pattern in rectangular samples. It contains
four 90$^\circ$ walls and one 180$^\circ$ domain wall in the
middle. The domain walls are assumed to be straight, thin lines. The
90$^\circ$ walls extend from the corners to the central 180$^\circ$
wall, which is parallel to the long edge of the rectangle and with
which they enclose an angle of 135$^\circ$.  To a very good
approximation, this pattern corresponds to experimental
observations of domain structures in rectangular magnetic platelets.
But also in particles of non-rectangular shape and even in extended
films \cite{Tsukahara72} this type of structure has been observed. 
In a hexagonal particle, the four-domain Landau in a rectangle
turns into a structure with six domains [cf.~Fig.\ref{VdB}b)]. A
distinction between this structure with six domains and the Landau
pattern may be neglected since the domain pattern is very similar
to the classical structure with four domains: two major domains and
two closure domains. Therefore, such modified variants also have been 
reasonably labeled  as ``Landau structure'', or sometimes as
``Landau-type structure'' in the literature, in spite of more or less
pronounced differences from the classical structure.  
In our case of thick particles, however, the differences are
significant and we require a different wording. Our study suggests
that in the three-dimensional 
case, the classical Landau structure is replaced by a complex flux
closure pattern with characteristic features. We shall call this the
{\em generalized} Landau structure and we will discuss this
structure in detail in
section \ref{internal}.

\section{Results}

\subsection{Asymmetric flux-closure patterns}
We have taken LEEM and XMCD-PEEM images of the magnetization at zero
field of about thirty Fe islands. In all cases, flux-closure
magnetization patterns have been observed.

\begin{figure}[h]
\centerline{\epsfig{file=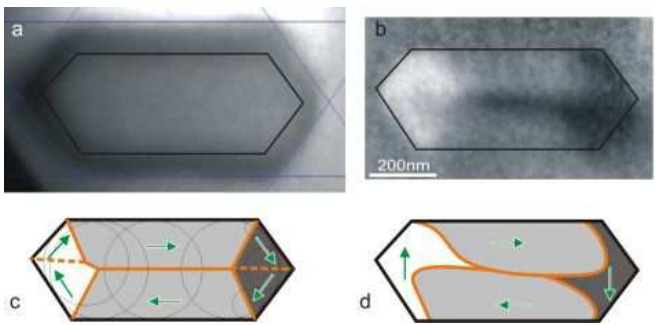,width=\linewidth}}
\caption{\label{noVdB}(color online) Difference between idealized and observed
  magnetic domain structure. a) LEEM image of the sample shape. 
  XMCD-PEEM image of the magnetic domain structure. 
  c) Schematic domain structure for the given shape according to the
  Van den Berg scheme. d) Approximate shape of the magnetic domains
  according to the experimental observation.}  
\end{figure}
Fig.~\ref{noVdB}b) shows an example of a typical domain
pattern observed in the experiments. The grey scale displays the $y$
component of the magnetization, {\em i.e.} the projection of the
magnetization at the surface along the polarization of the incident
beam.
The magnetic structure in fig.~\ref{noVdB}b) is clearly
subdivided into domains, but the shape of the domains is not in
accordance with the Van den Berg scheme, as can be seen by comparing
fig.~\ref{noVdB}c) and fig.~\ref{noVdB}d). The grey scale in these
schematic drawings refers again to the $y$ component, so that the
dashed domain walls sketched in Fig.~\ref{noVdB}c) would not be
visible in this experiment, since they separate regions of different
magnetization direction with the same value of $m_y$. Similar to the
classical Landau structure, the magnetization distribution in
Fig.~\ref{noVdB}b) is split into four domains: two major domains where
the magnetization is parallel to the long edge and two closure domains
that connect the magnetic flux between the major domains. However,
neither the shape of the domains nor the position of the domain walls
corresponds to the Van den Berg scheme.  

We have found such distorted, asymmetric domain patterns bearing only
a weak resemblance to the classical Landau structure in all the 
particles that we have investigated, as we will discuss in section 
\ref{compare_sec}.   
 The reason for the asymmetric structures and the deviations from the
 Van den Berg scheme lies in the three-dimensional nature of these
 particles. 

\subsection{Micromagnetic Simulations}

The XMCD-PEEM experiments only provide information about one
magnetization component (the component parallel to the incident beam)
on the flat, topmost surface of the sample. Therefore, numerical
simulation methods are employed to investigate the internal structure
of the magnetization. The simulation results are cross-checked with
the experimental observations to guarantee that the simulated
structures correspond to the observed ones. 

 Simulating the
magnetization in these particles is problematic because of their size,
which is enormous from the point of view of computational
micromagnetism. In particles of large size, it is difficult to avoid
discretization errors that result from too large discretization
cells. The consequence of discretization errors is not only a 
rough representation of the magnetization, but also wrong results
due to numerical ``domain wall collapse'' \cite{Donahue98}, {\em i.e.}~an unrealistic, 
large change of the magnetization direction within a single disretization 
cell. To avoid discretization errors, the size of the
discretization cells should not exceed the exchange length
$\Lambda=\sqrt{2A/(\mu_0M_s^2)}\simeq$ 3 nm. Since the particle volume is of
the order of 1$\mu{\rm m}^3$, simulations with a regular grid would
require an unsustainably large number of several millions of
discretization cells. An adaptive mesh refinement technique
\cite{Hertel98a} has been employed to solve this problem. 
Such adaptive methods are particularly powerful for
static micromagnetic simulations of large particles,
where the inhomogeneities of the magnetization are limited to
a small fraction of the sample's volume. These regions, {\em i.e.}~the
domain walls, require a high discretization density, whereas large 
domains with homogeneous magnetization can be calculated safely with a
coarse mesh. Therefore, adaptive refinement methods allow for 
accurate numerical results while keeping the computational costs
(processing time, memory) low.  

\begin{figure}
\centerline{\epsfig{file=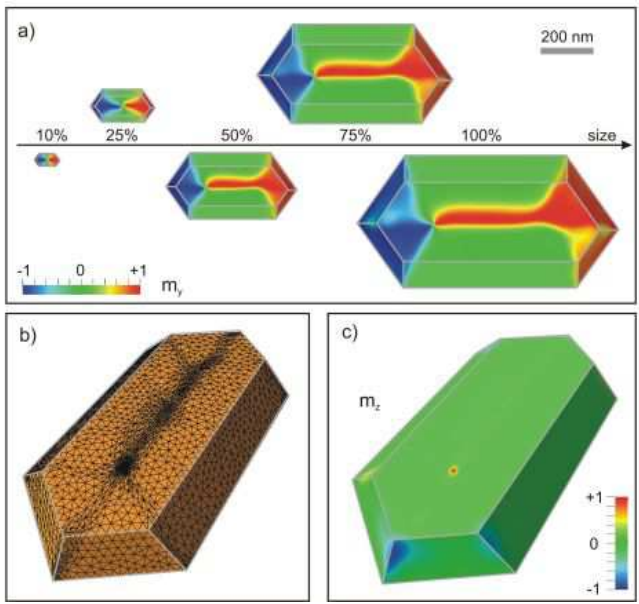,width=\linewidth}}
\caption{\label{inflate}(color online) To avoid discretization errors in the
  simulation of the three-dimensional magnetic domain structure, the
  sample is first reduced to a tenth of its 
  size. By using cycles of micromagnetic energy minimization and
  three-dimensional adaptive mesh refinement, the sample is
  gradually increased to its real size (a). After this procedure, a
  finite-element mesh with regions of strongly different
  discretization density is obtained (b). The finite elements are
  particularly small in regions of strong magnetic inhomogeneities
  (vortices, domain walls). Owing to the mesh refinement, even the
  nanometric vortex core with magnetization perpendicular to the
  surface is fully resolved, as can be seen in the red spot in
  panel (c). Throughout the paper, unless otherwise specified, the
  color coding used to represent a magnetization component is chosen
  according to the bar on the bottom of panel (a).}   
\end{figure}

The procedure we have applied for the micromagnetic simulations is as
follows. First, the exact geometrical shape of each Fe island is
extracted from LEEM images and a corresponding finite
element model is constructed. Subsequently, the model is scaled down
to 1/10 of its actual 
size. The magnetization distribution in this small island is
simulated by means of energy minimization, using a symmetric vortex
structure around the center as initial configuration. 
Once a converged solution is found, the mesh is adaptively refined and 
the equilibrium structure is calculated with the refined mesh. The
cycle of mesh refinement and energy minimization is repeated until the
maximum angle enclosed between the magnetization vectors at two neighboring
discretization points drops below $\pi/10$. Then, the size of the
sample is increased, and the procedure of mesh
refinement and energy minimization is repeated. With
this iterative process, the sample size is increased in small steps
towards its real value and the mesh is progressively refined. This way of
slowly ``inflating'' the sample guarantees that the magnetic structure
is calculated without discretization errors. The simpler variant of
simulating the magnetic structure directly with a coarse mesh, without
reducing the sample's size, and subsequently applying the  
mesh refinement procedure is less accurate because the first
calculation would lead to large discretization errors. Especially in
three-dimensional calculations, it is advisable to prevent
discretization errors rather than trying to remove them {\em a posteriori}.
An example for the simulation of the magnetization by means of  
stepwise increase of the sample size is shown in
Fig.~\ref{inflate}a). The typical cell size of a refined mesh is
between about 25 nm in the homogeneous regions and 1.5 nm in the
refined regions. 
By increasing the sample size, we find that the symmetric vortex
structure in small particles evolves to an asymmetric domain structure
above a certain sample size. Such a transition can be seen in
Fig.~\ref{inflate}a), where the asymmetry occurs after the sample has
been increased to 50\% of its actual size. The formation of the
asymmetric structure obviously involves a breaking of symmetry: the
vortex which is originally placed in the center may either shift to
the right or to the left side on the top surface. As will be
discussed later, the sign of the perpendicular component in the vortex
core may also have a strong influence on the resulting  domain
structure.   
\subsection{Comparison between Experiment and Simulation\label{compare_sec}}

\begin{figure*}
\centerline{\epsfig{file=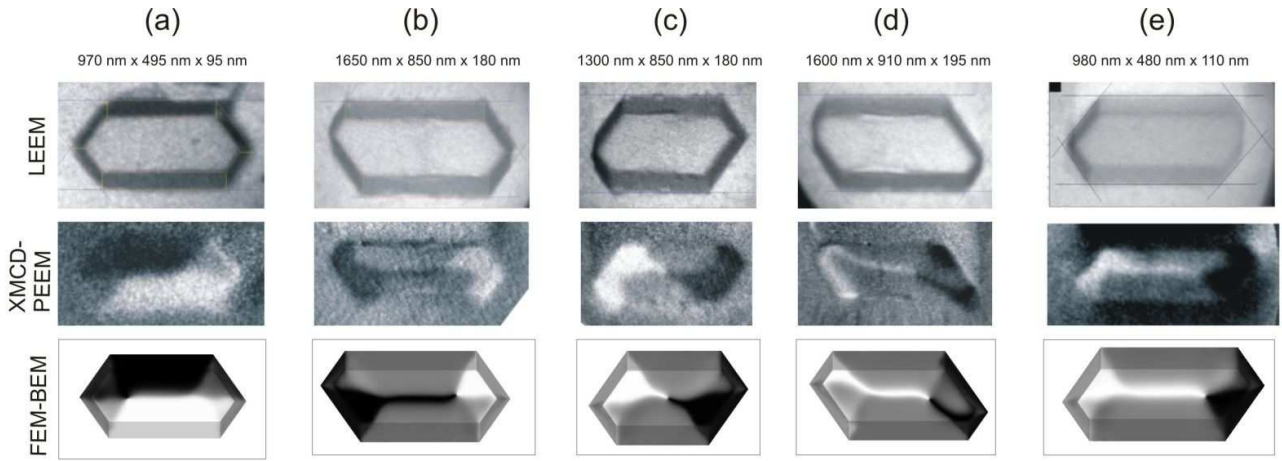,width=\textwidth}}
\caption{\label{compare}Experimental and simulated data for five
  self-assembled Fe/Mo(110) islands of different size and shape
  (a-e). {\em First row}: The shape of the
  particle is obtained by LEEM imaging. The particle thickness is
  derived from the width of the (001) facets, which appear as a dark border
  around the hexagonal top  surface, by using the known inclination
  angles of the inclined facets (cf.~Fig.~\ref{geom}). The sample's
  approximate lateral dimension and thickness is displayed on top of
  each column. {\em Second row:}  One magnetization component in the
  surface plane is imaged as grey 
  scale with  XMCD-PEEM. In sample (a), the contrast refers to the
  magnetization component parallel to the long edge while in the other
  samples the in-plane magnetization component perpendicular to the
  long edge is displayed. Notice that only the top surface, {\em i.e.}
  the light grey internal hexagon in the first row, is imaged. {\em
  Third row:} Micromagnetic simulation 
  results. To compare the results with the XMCD-PEEM experiments, the same
  magnetization component is displayed in grey scale. The sample shape
  was modelled according to the LEEM images and the model explained in
  Fig.~\ref{geom}. }
\end{figure*}

The simulation results are in excellent agreement with the
experimental data. A number of examples are shown in
Fig.~\ref{compare}. The experimentally observed asymmetric shape of
the end domains and the position of the domain walls are reproduced
almost perfectly by the simulations. All the unexpected features of the
domain patterns and their deviation from the Van den Berg scheme
occur in just the same way in both the experiment and the
simulations. Notice that all these details of the magnetic structure
develop automatically in the simulation by using a simple vortex as a
starting configuration.   

The one-to-one correspondence between experimental and simulated
surface magnetization is a strong indication that 
the simulated structures are indeed equal to the
experimental ones. Therefore, the simulations can be used  
to obtain information that is not directly accessible in the
experiment, {\em i.e.}~the three-dimensional structure of the magnetization
inside the sample. The investigation of the inner structure of the
magnetization is a prerequisite for the explanation of the
unexpected experimental results.

\section{Discussion}
\subsection{Internal domain wall strucure\label{internal}}
\begin{figure}

\centerline{\epsfig{file=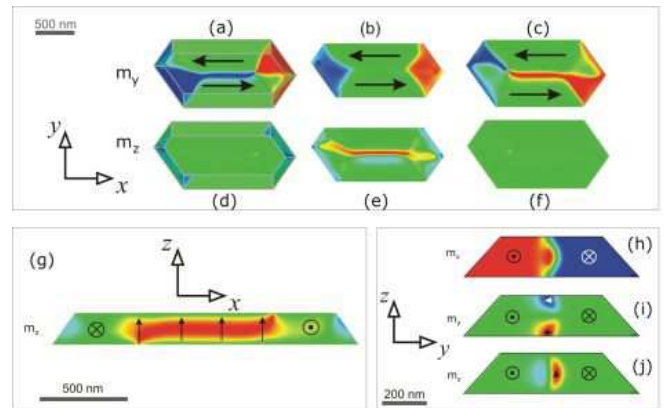,width=\linewidth}}

\caption{\label{cuts} (color online) Magnetization distribution on different
  cross sections. The example shown here refers to the particle
  displayed in Fig.~\ref{compare}b). The figures (a) and (d) show the top
  view on the three-dimensional sample with lateral facets. (b),(e):
  cross-section through the middle of the sample at $z=t/2$; (c),(f):
  Magnetization on the bottom surface of the sample, $z=0$. (g): $m_z$
  component on the cross-section through the middle of the sample
  parallel to the $xz$-plane. (h-j): asymmetric Bloch wall:
  magnetization components on a cross-section at mid-length, parallel
  to the $yz$-plane.}  
\end{figure}

\begin{figure}
\centerline{\epsfig{file=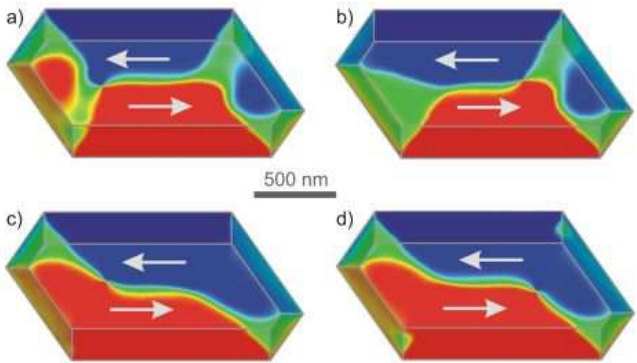,width=\linewidth}}
\caption{\label{isomers}(color online) Simulation of four different configurations in
  a sample of identical shape. The shape and the size of the sample is
  the same as in fig.~\ref{compare}d). Although topologically the
  structure is very similar in all cases (Landau structure with asymmetric Bloch wall),
  the resulting magnetic patterns are significantly different. 
  In the structures (a) and (b)  the Bloch component of the
  internal wall, $m_z$, is positive, whereas in (c) and (d) it is negative. The
  structures (a) and (c) have the same direction of the magnetization
  in the N\'eel caps, $m_y$, which is opposite to the orientation in the
  structures (b) and  (d).}    
\end{figure}

The magnetization distribution inside the sample differs significantly
from the  distribution observed at the surface. A typical example is
shown in Fig.~\ref{cuts}a),b,)c) that displays the $y$ component of the
magnetization on three cross-sections at the top ($z=t$), the middle
($z=t/2$) and the bottom ($z=0$), respectively. The coordinate frame
is chosen according to Fig.~\ref{geom}. In the middle cross section,
the domain pattern is almost symmetric and in good agreement with the
Van den Berg scheme, while on the top and the bottom the structure is 
clearly asymmetric. The twist is in  opposite directions on the top
and the bottom surface, cf.~\ref{cuts}a),c). The inhomogeneity of the 
magnetization along the $z$-direction and the 
resulting asymmetry in the top and bottom $xy$-plane  result from the
formation of an asymmetric 180$^\circ$ Bloch wall \cite{Hubert69,LaBonte69}: in the middle of
the sample ($z=t/2$) the major domains are separated by a Bloch wall,
while on the top and
bottom surface, the transition between the major domains is given by a
N\'eel wall (the so-called N\'eel caps). In a cross section on a plane with $x=$const.
through the middle of the sample, the typical profile of an asymmetric Bloch wall
resulting from the combination of Bloch and N\'eel wall can be clearly
seen \ref{cuts}h),i),j). 
There are different possibilities for the orientation of the
magnetization in this type of domain wall: The magnetization in the
Bloch part can point either in the positive 
or in the negative $z$ direction, and the magnetization in the N\'eel
caps can point either in the positive or negative $y$
direction. The orientation or the chirality of the asymmetric Bloch wall
can have a strong impact on the
resulting domain structure, as shown in Fig.~\ref{isomers}. Depending
on the direction of the magnetization of the N\'eel caps and the
direction of the inner Bloch component, the magnetic domain structure
on the surface may be very different.  In the example shown in
Fig.~\ref{isomers}, the computer simulation first yielded the
structure (a) for this geometry. In the simulations, the
aforementioned breaking of symmetry that is involved with the
formation of such structures results from numerical roundoff errors
and the direction in which the N\'eel caps evolve is
impredictable. This imponderability can be removed by breaking the
symmetry with an oblique, weak external field that is applied during
the simulated expansion of the sample, cf.~Fig.~\ref{inflate}a).  
With this controlled breaking of symmetry we obtained the structures 
shown in Fig.~\ref{isomers}, including the structure (d), which almost 
perfectly reproduces the experimental observation,
cf.~Fig.~\ref{compare}d). Apart from possible minor differences concerning
the out-of plane component of the magnetization along the facets,
there are four main types of such almost degenerate structures,
corresponding to the two possibilities for the orientation of the
Bloch component and the N\'eel caps, respectively \footnote{By flipping the
magnetization $\bm{M}\to-\bm{M}$, four further
domain structures can be obtained which correspond to a set of
magnetization states with opposite chirality of the central
vortex. Owing to time-inversion symmetry, the energy of these
structures is identical to the ones prior to the inversion of
$\bm{M}$, and their topology is obviously the same, too. It is
therefore sufficient to consider only one type of vortex chirality.}.
In a sense, these structures can be compared to geometric isomers of a 
molecule: The domain structures contain the same
components (Landau structure with asymmetric
internal walls), but the orientation is different (opposite orientation
of the N\'eel caps and/or of the Bloch wall).  The domain structures
of such ``isomers'' can differ significantly in their patterns on the
surface, especially if the sample shape is strongly asymmetric. 

In any case, the N\'eel caps of an asymetric Bloch wall are
oriented in opposite direction on opposite surfaces. The formation of
an asymmetric Bloch wall represents an energetically optimized
arrangement concerning the stray field energy. Magnetostatic volume charges
$\rho=\bm{\nabla}\cdot\bm{ M}$  are largely avoided by the formation of a Bloch
wall in the center of the sample and surface charges
$\sigma=\bm{M}\cdot\bm{\hat{n}}$ ($\bm{\hat{n}}$: surface normal vector)
are effectively suppressed by the N\'eel caps.  
Depending on whether the magnetization in the N\'eel caps is pointing
parallel or antiparallel to the magnetization direction in the closure
domains (the domains close to the particle's border in the $\pm x$
direction), the N\'eel caps are either connected smoothly to the closure domain 
or separated by them with a magnetic vortex, respectively \cite{Arrott97}.
Inside the closure domain, on  cross-sections  with constant value of
$x$, the structure is similar to an asymmetric N\'eel wall. The main
difference between an asymmetric N\'eel wall and an asymmetric Bloch wall
is the relative orientation of the magnetization in the N\'eel caps,
being parallel in the former and antiparallel in the latter.
It has been predicted, by means of numerical simulations, 
that the combination of asymmetric wall types in mesoscopic,
soft-magnetic particles leads to a pronounced in-plane asymmetry on
the top surfaces \cite{Hertel99a,Hertel99b}. Our XMCD-PEEM
investigations on several mesoscopic Fe islands clearly confirm this
prediction and provide the first direct observation of this effect.

\subsection {Shape dependence\label{generalized}}
In the numerous particles that we have investigated, we have almost
exclusively found the Landau-type flux-closure pattern described
above, the {\em generalized} Landau structure. A few
exceptions of different magnetization arrangements will 
be discussed in the next sections. The generalized Landau structure
consists of four domains, two major domains that are  
connected by an asymmetric Bloch wall and two closure
domains connected to the major domains by 90$^\circ$ walls. 
Our observation of
the Landau structure in self-assembled Fe islands with their quite
complicated hexagonal shape and the inclined facets suggests that the
formation of this arrangement of the magnetization is rather driven by
the particle's aspect ratio and its size than by the rectangular shape. 

\begin{figure}
\centerline{\epsfig{file=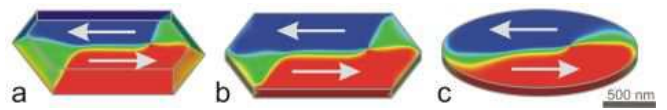,width=\linewidth}}
\caption{\label{cookie}(color online) Geometry dependence of the magnetization
  structure in a three-dimensional island. (a): Perspective view
  on the simulated  magnetization structure of the particle shown in
  Fig.~\ref{compare}b). (b): Simulated magnetic domain
  structure in a particle of same size and thichkness, but with
  symmetric in-plane shape and vertical facets. (c): Even if the
  hexagonal shape is replaced by a smooth elliptical boundary, the
  magnetization structures remains almost unchanged, with the same
  asymmetric characteristics.} 
\end{figure}

The question arises to which extent the asymmetric Landau pattern that we
observed is specific to our example of hexagonal particles with
inclined facets. How sensitively does the magnetic structure depend on the
precise shape of the particle? To answer this question, the influence
of the particle shape on the resulting magnetic domain structure has
been investigated with micromagnetic simulations. Unlike the real
world, where the shape of the Fe islands is determined by the
energetics of the growth mode, finite element
models give the possibility to study  magnetic structures in
mesoscopic particles of arbitrary shape.  Starting from a ``real''
island, {\em i.e.}~one that is modelled exactly  according to the
experiment, we have first tested to which extent the 
magnetic pattern is affected by the inclined facets and the asymmetry
of the hexagon. This has been done by simulating the  
magnetization in a hexagonal particle of same thickness and size, but
with symmetric hexagonal shape and a ``cookie-cutter'' geometry,
{\em i.e.}~an island with perpendicular facets.  The result is displayed in
Fig.~\ref{cookie}b). The asymmetric magnetic structure is practically the
same as it has been observed in the real island with inclined facets,
indicating that neither the inclination of the facets nor the asymmetry of the 
hexagon has a decisive impact on the magnetic structure. To further check
whether the hexagonal shape is of importance for the resulting magnetic
structure, a flat, elliptical particle has been modelled. The major axes
of the ellipse (parallel to the $x$ and $y$ direction, respectively) 
correspond to the width and length of the original island and the
thickness is again the same. The result [Fig.~\ref{cookie}c)] shows that
the distorted Landau structure also occurs in this
particle. Recent studies \footnote{R. Hertel
  {\em et al.}, unpublished.} show 
that practically the same structure may also be found
in an ellipsoidal particle of comparable size and aspect
ratio.
These computer experiments indicates that the generalized Landau
structure is hardly determined by the particle shape. This is in
contrast to two-dimensional samples, where the essential importance of
the sample shape reflects in the Van den Berg construction scheme,
where the domain structure is determined {\em exclusively} by means of
the particle shape. The occurrence of the three-dimensional,
generalized Landau structure is neither restricted to particles of
rectangular, hexagonal or any other specific shape; not even
to flat particles.  
In the three-dimensional case, it is rather the thickness, the size and
the aspect ratio that are decisive. The primary tendency of the
magnetization in thick, three-dimensional particles appears to be to
develop an asymmetric 180$^\circ$ Bloch wall, while in the case of
thin-film elements, the major domains may either be separated by a
180$^\circ$ wall (classical Landau structure) or by 90$^\circ$ walls 
(diamond state).

\subsection{Diamond structure}
By far, the magnetization states that we have observed most frequently are the 
above-mentioned variants of the Landau structure. The absence of the
``diamond state'' in these thick elements is remarkable. The diamond
state is an alternative flux-closure pattern  with  seven domains
(cf.~Fig.\ref{VdB}) that is frequently observed in thinner
soft-magnetic particles \cite{Jubert03,Fruchart04,Hertel99a} of
similar size.  

\begin{figure}
\centerline{\epsfig{file=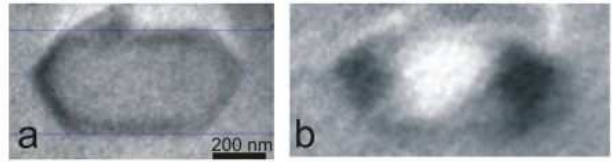,width=\linewidth}}
\caption{\label{Diamond}In a relatively large and particularly thin
island, identifiable as such in a LEEM image (a) by the thin dark
boundary (the lateral facets), a ``diamond state'' magnetization
structure was found. The diamond state is characterized by two 
vortices with opposite vorticity and a diamond-shaped central 
domain, which is clearly recognizable on the right panel (b).}
\end{figure}
We suspect that the predominance of the Landau
structure and the absence of the diamond state can be explained with
energetic considerations in connection with the elevated particle
thickness. The thickness dependence of domain
configurations in nanoscale Fe/W(110) islands of significantly lower
thickness (ca. 3-8 nm) has been studied recently by Bode {\em et al.}
\cite{Bode04}.
In the present case, the idea is the
following. Essentially, the Landau 
structure has one central 180$^\circ$ wall and four 90$^\circ$ domain
walls, whereas the diamond structure consists of seven domains that
are separated by {\em eight} 90$^\circ$ domain walls. While in the
case of a 180$^\circ$ wall magnetic volume charges can efficiently be
avoided by the formation of an asymmetric Bloch wall, a 90$^\circ$
wall inevitably contains volume charges $\rho=\bm{\nabla M}$, like a
N\'eel wall does. With increasing thickness, the energy connected with
these volume charges increases and so does the tendency to avoid
them. The tendency to avoid these charged 90$^\circ$ walls and,
instead, to form a 180$^\circ$ wall, favors the 
formation of the Landau structure over the diamond structure.  If this
consideration is correct, the tendency to avoid the diamond state
should decrease with decreasing thickness. We therefore searched for
the diamond state in islands of large in-plane extension (to ensure
that the particle is in a multi-domain state) that were particularly
thin. In our set of samples the typical thickness was about 100-150~nm, but we
eventually found an island of 65 nm thickness and about 1000 nm
$\times$ 500 nm lateral extension. The XMCD-PEEM investigation of the
magnetization in this island indeed showed that the sample was
magnetized in the diamond state, cf.~Fig.~\ref{Diamond}. This sample
is the only one in which we found the diamond state. The fact that we
have observed this state only after a selective search aimed at
finding particularly thin elements suggests that the thickness,
actually, plays a decisive role. Further studies are required to
systematically analyze this suggested thickness dependence of magnetization states.

\subsection{Vortex structure}
Besides the thickness and the size, also the in-plane aspect ratio has
an impact on the resulting magnetization state. This can be seen by
comparing the magnetic structure of two islands of similar thickness and
width, but different length,  cf.~Fig.~\ref{compare})b),c). In one
case, the aspect ratio (length in $x$ direction vs. width in $y$
direction) is 2.1 and in the other it is about 1.5. The particle with
lower aspect ratio has a vortex in the center of the sample. It
contains no asymmetric Bloch wall and the closure domains 
are asymmetric and twisted. They are stretched in the $x$
direction and they join in the middle of the sample at the vortex. In
contrast to this, the longer sample has a clearly developed
180$^\circ$ domain wall in the middle which separates the major
domains. Obviously, a sufficiently large
particle with aspect ratio approximately equal to one would not form a
Landau structure with a central 180$^\circ$ domain wall, but would be 
magnetized in a simple vortex state with a magnetic vortex structure in
the centre of the sample. With increasing aspect ratio, the vortex
core is stretched until it eventually performs a kink through the
sample and forms the center of an 
asymmetric Bloch wall. Thus, by changing the aspect ratio, a
transition from a vortex state to a generalized Landau structure
occurs at a critical size.  
Whether this transition is continuous or
discontinuous may be the subject of future studies.

\subsection{Bloch switch}

\begin{figure}
\centerline{\epsfig{file=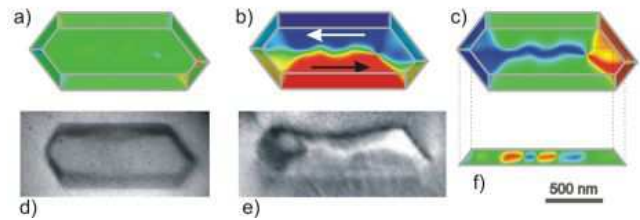,width=\linewidth}}
\caption{\label{wave}(color online) (a-c): $z, x$ and $y$ component of the
magnetization of a simulated magnetic domain structure with undulated
central domain wall. (f) $z$-component of the magnetization on a
cross-section of constant $y$. The alternating red-blue
regions indicate switches of the Bloch component.  (d): LEEM image and
XMCD-PEEM image (e) of a sample in an apparently similar magnetization
state. The size of the island is 1500 nm $\times$ 620 nm $\times$ 130 nm. }
\end{figure}

In most cases the simulations yield  magnetic
structures that agree very well with the ones that have been observed in the
experiments, but occasionally some interesting differences between
simulation and experiments have occurred. 
The possible differences concerning the chirality of the asymmetric
central Bloch wall have already been discussed in section
\ref{internal}. 
Another variant of the Landau structure was obtained in one case,
where the simulation has yielded a structure as shown in
Fig.~\ref{wave}b), c). Instead of the usual straight line, the central
domain wall ({\em i.e.}~the N\'eel cap) displays a pronounced
kink. It is known from studies of asymmetric domain walls in extended
magnetic films that such a kink is characteristic for a so-called
Bloch switch \cite{Harrison73,Hubert98}. A Bloch switch is given when
the sign of the Bloch component changes
along the domain wall inside the film,  while the magnetization
direction of the N\'eel caps remains unchanged. In fact, a look at the
inner part of the simulated magnetization structure
[Fig.~\ref{wave}f)]
shows that the arrangement contains Bloch switches. We have also observed a
similar kink in the experiment, as shown in Fig.~\ref{wave}e). 

This example is the only island in
which we have found this kink experimentally, and it can therefore
not be ruled out that the kink may have other reasons than a
Bloch switch like, {\em e.g.}, a structural inhomogeneity of the
island. Moreover, the top surface of
the island in which we observed the kink was not ideally flat, as
could be seen by the shadow of the island as described in
Fig.~\ref{example}b). Due to these deviations of the sample morphology
from the ideal construction scheme according to Fig.~\ref{geom}, the
model for the island shape that has been used in the simulations that
yielded a Bloch switch does not correspond exactly to the island in
which the kink has been observed experimentally. Therefore, although a
correspondence between experiment and simulation is likely, the
available data from simulation and experiment are not sufficient for
an unambiguous proof that the kink results from a Bloch switch. The
results, however, suggest that an asymmetric Bloch wall with a Bloch
switch may develop also in patterned particles and that such a Bloch
switch would lead to a considerable distortion of the otherwise
straight central domain wall. To our knowledge, this kind of domain
wall has not been reported previously in patterned elements.

\section{Conclusions}
``The simplest problem in continuum magnetism in a singly connected
finite body is the Landau structure (\dots), but it remains an unsolved
problem to describe it in all its details.'' This is a quote from a
paper by A.S. Arrott and T.L. Templeton \cite{Arrott97} that was
written in 1997, at a  time when the experimental resolution was not
yet sufficient for a direct observation of details of the
magnetization and most micromagnetic simulation methods were not
accurate enough for large, three-dimensional computations \footnote{
  Similarly, Aharoni stated in his 
  textbook \cite{Aharoni96_2nd} (p.173): ``With present computers
  it is not even possible to find the lowest-energy configuration of a
  single wall,(\dots) let alone (\dots) any (\dots) true three-dimensional
  magnetization distribution''.}.
Owing to the combination of 
(a) advanced growth methods yielding high-quality, monocrystalline
mesoscopic Fe particles that can be regarded as model systems, 
(b) LEEM imaging to obtain detailed information on the particle shape,
(c) high-resolution XMCD-PEEM studies that provide information on the
in-plane manetization at the particle surface, and (d) accurate
FEM/BEM micromagnetic modelling with adaptive mesh refinement
techniques providing a detailed model of the three-dimensional
magnetization structure, we have solved the problem of a full
description of the Landau
structure. The Landau structure seems to occur as a minimum energy
arrangement in soft-magnetic particles of very different
shape. Generally, the size, thickness and aspect ratio are more
important for the formation of the generalized Landau structure than
details of the particle's shape and its surfaces.  The fascinating
complexity of the Landau structure occurs when the particle is thick
enough to sustain an asymmetric Bloch wall. The classical, symmetric
Landau pattern occurring in rectangular thin film elements can be
regarded as a simple thin-film limit of a more general, three-dimensional
structure. In the thin-film limit, the central asymmetric Bloch wall is
replaced by an ordinary N\'eel walls, thus simplifying drastically the
arrangement.

\section*{Acknowledgements}
We acknowledge financial support from the EU VI Framework Program
Transnational Access [RII3-CT-2004-506008(IA-SFS)] and Laboratoire
Europ\'een Associ\'e (LEA) Mesomag (MPI-Halle/LLN). We thank
V. Santonacci and Ph.~David for technical support on the UHV growth
chambers. We also acknowledge support by the European Community -
Research Infrastructure Action under the FP6 ``Structuring the European
Research Area'' Programme (through the Integrated Infrastructure
Initiative ``Integrating Activity on Synchrotron and Free Electron
Laser Science'')  

\bibliography{/tmp_mnt/users/iff_ee/hertel/bib/Literatur.bib}

\begin{thebibliography}{10}

\bibitem{Martin03}
J.~I. Mart{\'i}n, J.~Nogu{\'e}s, K.~Liu, J.~L. Vicent, and I.~K. Schuller,
\newblock Ordered magnetic nanostructures: fabrication and properties,
\newblock J. Magn. Magn. Mater. {\bf 256}, 449--501 (2003).

\bibitem{Cowburn99}
R.~P. Cowburn, D.~K. Koltsov, A.~O. Adeyeye, M.~E. Welland, and D.~M. Tricker,
\newblock Single-Domain Circular Nanomagnets,
\newblock Phys. Rev. Lett. {\bf 83}, 1042--1045 (1999).

\bibitem{Kirk97}
K.~J. Kirk, J.~N. Chapman, and C.~D.~W. Wilkinson,
\newblock Switching fields and magnetostatic interactions of thin film magnetic
  nanoelements,
\newblock Appl. Phys. Lett. {\bf 71}(4), 539--541 (1997).

\bibitem{Gomez99}
R.~D. Gomez, T.~Luu, A.~Pak, I.~Mayergoyz, K.~Kirk, and J.~Chapman,
\newblock Domain wall motion in micron sized Permalloy elements,
\newblock J. Appl. Phys. {\bf 85}(8), 4598--4600 (1999).

\bibitem{Hubert98}
A.~Hubert and R.~Sch\"afer,
\newblock {\em Magnetic Domains - The Analysis of Magnetic Microstructures},
\newblock Berlin, New York, Heidelberg: Springer, 1998.

\bibitem{Brown63}
W.~F. Brown, Jr.,
\newblock {\em Micromagnetics},
\newblock Interscience Publishers, John Wiley \& Sons, New York, London, 1963.

\bibitem{Aharoni96_2nd}
A.~Aharoni,
\newblock {\em Introduction to the Theory of Ferromagnetism},
\newblock Oxford Science Publications, Oxford Clarendon Press, Oxford, 2nd.
  edition, 1996, 2000.

\bibitem{Stoner48}
E.~C. Stoner and E.~P. Wohlfarth,
\newblock A Mechanism of magnetic hysteresis in heterogeneous alloys,
\newblock Phil.Trans.Roy.Soc. {\bf 240}, 599 (1948).

\bibitem{Wernsdorfer97}
W.~Wernsdorfer, E.~B. Orozco, K.~Hasselbach, A.~Benoit, B.~Barbara, N.~Demoncy,
  A.~Loiseau, H.~Pascard, and D.~Mailly,
\newblock Experimental evidence of the N\'eel-Brown model of magnetization
  reversal,
\newblock Phys. Rev. Lett. {\bf 78}(9), 1791--1794 (1997).

\bibitem{Yamasaki03}
A.~Yamasaki, W.~Wulfhekel, R.~Hertel, S.~Suga, and J.~Kirschner,
\newblock Direct Observation of the Single-domain Limit of Fe nano-Magnets by
  Spin-Polarized STM,
\newblock Phys. Rev. Lett. {\bf 91}(12), 1272011 (2003).

\bibitem{Hertel02c}
R.~Hertel,
\newblock Thickness dependence of magnetization structures in thin Permalloy
  rectangles,
\newblock Z. Metallkd. {\bf 93}(10), 957--962 (2002).

\bibitem{Usov01}
N.~A. Usov, C.~R. Chang, and Z.~H. Wei,
\newblock Nonuniform magnetization structures in thin soft type ferromagnetic
  elements of elliptical shape,
\newblock J. Appl. Phys. {\bf 89}(11), 7591--7593 (2001).

\bibitem{Kroni62}
H.~Kronm{\"u}ller,
\newblock Mikromagnetische Berechnung der Magnetisierung in der Umgebung
  unmagnetischer Einschl{\"u}sse in Ferromagnetika,
\newblock Zeitschrift f\"ur Physik {\bf 168}, 478--494 (1962).

\bibitem{Rothman01}
J.~Rothman, M.~Kl\"aui, L.~Lopez-Diaz, C.~A.~F. Vaz, A.~Bleloch, J.~A.~C.
  Bland, Z.~Cui, and R.~Speaks,
\newblock Observation of a Bi-Domain State and Nucleation Free Switching in
  Mesoscopic Ring Magnets,
\newblock Phys. Rev. Lett. {\bf 86}(6), 1098--1101 (2001).

\bibitem{Nielsch2001}
K.~Nielsch, R.~B. Wehrspohn, J.~Barthel, J.~Kirschner, U.~G\"osele, S.~F.
  Fischer, and H.~Kronm\"uller,
\newblock Perpendicular storage medium based on 100 nm period Nickel nanowire
  arrays,
\newblock Appl. Phys. Lett. {\bf 79}, 1360 (2001).

\bibitem{Park03}
M.~H. Park, Y.~K. Hong, S.~H. Gee, D.~W. Erickson, and B.~C. Choi,
\newblock Magnetization configuration and switching behavior of submicron NiFe
  elements: Pac-man shape,
\newblock Appl. Phys. Lett. {\bf 83}(2), 329--331 (2003).

\bibitem{Kirk01}
K.~J. Kirk, S.~McVitie, J.~N. Chapman, and C.~D.~W. Wilkinson,
\newblock Imaging magnetic domain structure in sub-500 nm thin film elements,
\newblock J. Appl. Phys. {\bf 89}(11), 7174--7176 (2001).

\bibitem{Hertel99b}
R.~Hertel and H.~Kronm\"uller,
\newblock Computation of the magnetic domain structure in bulk Permalloy,
\newblock Phys. Rev. B {\bf 60}(10), 7366--7378 (1999).

\bibitem{Rave98}
W.~Rave, K.~Fabian, and A.~Hubert,
\newblock Magnetic states of small cubic particles with uniaxial anisotropy,
\newblock J. Magn. Magn. Mater. {\bf 190}, 332--348 (1998).

\bibitem{Ha03a}
J.~K. Ha, R.~Hertel, and J.~Kirschner,
\newblock Micromagnetic study of magnetic configurations in submicron permalloy
  disks,
\newblock Phys. Rev. B {\bf 67}, 224432 (2003).

\bibitem{Hehn96}
M.~Hehn, K.~Ounadjela, J.-P. Bucher, F.~Rousseaux, D.~Decanini, B.~Bartenlian,
  and C.~Chappert,
\newblock Nanoscale magnetic domains in mesoscopic magnets,
\newblock Science {\bf 272}, 1782--1785 (1996).

\bibitem{Jubert03}
P.~O. Jubert, J.~C. Toussaint, O.~Fruchart, C.~Meyer, and Y.~Samson,
\newblock Flux-closure-domain states and demagnetizing energy determination in
  sub-micron size magnetic dots,
\newblock Europhys. Lett. {\bf 63}(1), 132--138 (2003).

\bibitem{Bloch32}
F.~Bloch,
\newblock Zur Theorie des Austauschproblems und der Remanenzerscheinung der
  Ferromagnetika,
\newblock Z. Physik {\bf 74}, 295--335 (1932).

\bibitem{Neel55}
L.~N\'eel,
\newblock Energies des parois de Bloch dans les couches minces,
\newblock C. R. Acad. Sci. Paris {\bf 241}, 533--536 (1955).

\bibitem{Moon59}
R.~M. Moon,
\newblock Internal Structure of Cross-Tie Walls in Thin Permalloy Films through
  High-Resolution Bitter Techniques,
\newblock J. Appl. Phys. {\bf 30}(4), 82S--83S (1959).

\bibitem{Hubert69}
A.~Hubert,
\newblock Stray-field-free magnetization configurations,
\newblock phys. stat. sol. {\bf 32}(2), 519 (1969).

\bibitem{Jubert04}
P.~O. Jubert, R.~Allenspach, and A.~Bischof,
\newblock Magnetic domain walls in constrained geometries,
\newblock Phys. Rev. B {\bf 69}(22), 220410 (2004).

\bibitem{Bruno99}
P.~Bruno,
\newblock Geometrically Constrained Magnetic Wall,
\newblock Phys. Rev. Lett. {\bf 83}(12), 2425--2428 (1999).

\bibitem{McMichael97b}
R.~D. McMichael and M.~J. Donahue,
\newblock Head to head domain wall structures in thin magnetic strips,
\newblock IEEE Trans. Magn. {\bf 33}(5), 4167--4169 (1997).

\bibitem{Klaui04}
M.~Kl\"aui, C.~A.~F. Vaz, J.~A.~C. Bland, L.~J. Heyderman, F.~Nolting,
  A.~Pavlovska, E.~Bauer, S.~Cherifi, S.~Heun, and A.~Locatelli,
\newblock Head-to-head domain-wall phase diagram in mesoscopic ring magnets,
\newblock Appl. Phys. Lett. {\bf 85}(23), 5637--5639 (2004).

\bibitem{Fruchart98}
O.~Fruchart, S.~Jaren, and J.~Rothman,
\newblock Growth modes of W and Mo thin epitaxial (110) films on (11$\bar{2}$0)
  Sapphire,
\newblock Appl. Surf. Sci. {\bf 135}, 218--232 (1998).

\bibitem{Jubert01}
P.~O. Jubert, O.~Fruchart, and C.~Meyer,
\newblock Self-assembled growth of faceted epitaxial Fe(110) islands on
  Mo(110)/Al$_2$0$_3$(11$\bar{2}$0),
\newblock Phys. Rev. B {\bf 64}, 115419 (2001).

\bibitem{Fruchart04}
O.~Fruchart, J.-C. Toussaint, P.-O. Jubert, W.~Wernsdorfer, R.~Hertel,
  J.~Kirschner, and D.~Mailly,
\newblock Angular dependence of magnetization switching for a multidomain dot:
  Experiment and simulation,
\newblock Phys. Rev. B {\bf 70}, 172409 (2004).

\bibitem{Mueller00}
P.~M\"uller and R.~Kern,
\newblock Equilibrium nano-shape changes induced by epitaxial stress
  (generalised Wulf-Kaishew theorem),
\newblock Surf. Sci. {\bf 457}(1-2), 229--253 (2000).

\bibitem{Stoehr93}
J.~St\"ohr, B.~D. Hermsmeier, M.~G. Samant, G.~R. Harp, S.~Koranda, D.~Dunham,
  and B.~P. Tonner,
\newblock Element-specific magnetic microscopy with circularly polarized
  x-rays,
\newblock Science {\bf 259}, 658--661 (1993).

\bibitem{Locatelli02}
A.~Locatelli, S.~Cherifi, S.~Heun, M.~Marsi, K.~Ono, A.~Pavlovska, and
  E.~Bauer,
\newblock X-ray magnetic circular dichroism imaging in a low energy electron
  microscope,
\newblock Surface Review and Letters {\bf 9}(1), 171--176 (2002).

\bibitem{Locatelli03}
A.~Locatelli, A.~Bianco, D.~Cocco, S.~Cherifi, S.~Heun, M.~Marsi,
  M.~Pasqualetto, and E.~Bauer,
\newblock High lateral resolution spectroscopic imaging of surfaces: The
  undulator beamline "nanospectroscopy" at Elettra,
\newblock Journal de Physique IV {\bf 104}, 99--102 (2003).

\bibitem{Hertel01b}
R.~Hertel,
\newblock Micromagnetic simulations of magnetostatically coupled Nickel
  nanowires,
\newblock J. Appl. Phys. {\bf 90}(11), 5752--5758 (2001).

\bibitem{NumRec}
W.~H. Press, B.~P. Flannery, S.~A. Teukolsky, and W.~T. Vetterling,
\newblock {\em Numerical Recipies: The art of scientific computing},
\newblock Cambrige University Press, Cambrige, New York, 1986.

\bibitem{VdBerg86}
H.~A.~M. Van Den~Berg,
\newblock Self-consistent domain theory in soft-ferromagnetic media II. Basic
  domain structures in thin film objects,
\newblock J. Appl. Phys. {\bf 60}, 1104--1113 (1986).

\bibitem{VdBerg87}
H.~A.~M. Van Den~Berg and A.~H.~J. Van Den~Brandt,
\newblock Self-consistent domain theory in soft-ferromagnetic media III.
  Composite domain structures in thin film objects,
\newblock J. Appl. Phys. {\bf 62}, 1952--1959 (1987).

\bibitem{Rave00}
W.~Rave and A.~Hubert,
\newblock Magnetic ground state of a thin-film element,
\newblock IEEE Trans. Magn. {\bf 36}(6), 3886--3889 (2000).

\bibitem{Arrott79}
A.~S. Arrott, B.~Heinrich, and A.~Aharoni,
\newblock Point Singularities and Magnetization Reversal in Ideally Soft
  Ferromagnetic Cylinders,
\newblock IEEE Trans. Magn. {\bf 15}(5), 1228--1235 (1979).

\bibitem{Arrott79b}
A.~S. Arrott and J.-G. Lee,
\newblock Simple function for a complex domain configuration,
\newblock J. Appl. Phys. {\bf 79}(8), 5752--5754 (1979).

\bibitem{Landau35}
L.~Landau and E.~M. Lifshitz,
\newblock On the theory of the dispersion of magnetic permeability in
  ferromagnetic bodies,
\newblock Phys. Z. Sowjet. {\bf 8}, 153 (1935).

\bibitem{Tsukahara72}
S.~Tsukahara and H.~Kawakatsu,
\newblock Magnetic Contrast and Inclined 90$^\circ$ Domain Walls in Thick Iron
  Films,
\newblock J.~Phys.~Soc.~Japan {\bf 32}(1), 72--78 (1972).

\bibitem{Donahue98}
M.~J. Donahue,
\newblock A variational approach to exchange energy calculations in
  micromagnetics,
\newblock J. Appl. Phys. {\bf 83}, 6491--6493 (1998).

\bibitem{Hertel98a}
R.~Hertel and H.~Kronm\"uller,
\newblock Adaptive Finite Element Mesh Refinement Techniques in
  Three-Dimensional Micromagnetic Modeling,
\newblock IEEE Trans. Magn. {\bf 34}(6), 3922--3930 (1998).

\bibitem{LaBonte69}
A.~E. LaBonte,
\newblock Two-Dimensional Bloch-Type Domain Walls in Ferromagnetic Films,
\newblock J. Appl. Phys. {\bf 40}(6), 2450--2458 (1969).

\bibitem{Arrott97}
A.~S. Arrott and T.~L. Templeton,
\newblock Micromagnetics and the hysteresis as prototypes for complex systems,
\newblock Physica B {\bf 233}(4), 259--271 (1997).

\bibitem{Hertel99a}
R.~Hertel and H.~Kronm\"uller,
\newblock Micromagnetic simulation of the domain structure of a flat
  rectangular permalloy prism,
\newblock J. Appl. Phys. {\bf 85}(8), 6190--6192 (1999).

\bibitem{Bode04}
M.~Bode, A.~Wachowiak, J.~Wiebe, A.~Kubetzka, M.~Morgenstern, and
  R.~Wiesendanger,
\newblock Thickness dependent magnetization states of Fe islands on W(110):
  From single-domain to vortex and diamond patterns,
\newblock Appl. Phys. Lett. {\bf 84}(6), 948--950 (2004).

\bibitem{Harrison73}
C.~G. Harrison and K.~D. Leaver,
\newblock The Analysis of Two-Dimensional Domain Wall Structures by Lorentz
  Microscopy,
\newblock phys.~stat.~sol.(a) {\bf 15}, 415--429 (1973).

\end{thebibliography}
\bibliographystyle{phreport}
\end{document}